\documentclass
[aps,preprint,12pt,showpacs,showkeys,byrevtex,superbib,eqsecnum,fleqn]{revtex4}%
\usepackage{amsfonts}
\usepackage{longtable}
\usepackage{amsmath}
\usepackage{amssymb}
\usepackage{graphicx}
\usepackage{acronym}
\usepackage{amsbsy}
\usepackage{amstext}%
\newtheorem{theorem}{Theorem}

\newtheorem{proposition}[theorem]{Proposition}

\newenvironment{proof}[1][Proof]{\noindent\textbf{#1.} }{\ \rule{0.5em}{0.5em}}
\begin{document}
\preprint{HEP/123-qed}
\title[Short title for running header]{Isospectral Hamiltonian for position-dependent mass for an arbitrary quantum
system and coherent states}
\author{Sid-Ahmed Yahiaoui and Mustapha Bentaiba}
\email{bentaiba@univ-blida.dz}
\affiliation{LPTHIRM, d\'{e}partement de physique, facult\'{e} des sciences, universit\'{e}
Sa\^{a}d DAHLAB-Blida1, B.P. 270 Route de Soum\^{a}a, 09000 Blida, Algeria}

\begin{abstract}
By means of the unitary transformation, a new way for discussing the ordering
prescription of Schr\"{o}dinger equation with a position-dependent mass (PDM)
for isospectral Hamiltonian operators is presented. We show that the ambiguity
parameter choices in the kinetic part of the Hamiltonian can be explained
through an exact SUSY symmetry as well as a consequence of an accidental
symmetry under the $%
\mathbb{Z}
_{2}$ action. By making use of the unitary transformation, we construct
coherent states for a family of PDM isospectral Hamiltonians from a suitable
choice of ladder operators. We show that these states preserve the usual
structure of Klauder-Perelomov's states and thus saturate and minimize the
generalized position-momentum uncertainty relation (GUR) under some special
restrictions. We show that GUR's properties can be used to determine the sign
of the superpotential.

\end{abstract}
\volumeyear{year}
\volumenumber{number}
\issuenumber{number}
\eid{identifier}
\date[Date text]{date}
\received[Received text]{date}

\revised[Revised text]{date}

\accepted[Accepted text]{date}

\published[Published text]{date}

\startpage{1}
\endpage{ }
\maketitle

\section{Introduction}

Nowadays, it is well-known that a family of Hamiltonians with the same
eigenvalues set of the original Hamiltonian is called isospectral Hamiltonians
\cite{1,2}. These families have been known for a long time and were obtained
by means of factorization method \cite{3}, the Gel'fand-Levitan's approach
\cite{4}, supersymmetric quantum mechanics (SUSY) and the related
shape-invariant \cite{5} and Lie algebraic procedure \cite{6}. These
techniques are known as a most fruitful approaches for solving Schr\"{o}dinger
equation in the context of constant mass.

When working with quantum system subjected to interact with a given
interaction, however, usually considerations impose to identify the mass-term
with the concept of effective mass. By a way, such quantum system becomes
position-dependent mass (PDM) and considerable interest has been recently
devoted in studying the Schr\"{o}dinger equation under this new perspective
\cite{7,8,9,10,11,12,13,14,15,16,17}. We find the most extensive use of such
an equation in effective interactions in nuclear physics \cite{18}, in the
framework of curved spaces \cite{19}, in the point of view of $\mathcal{PT}%
$-symmetry \cite{20,21}, quantum wells, wires and dots \cite{22} and
semiconductor heterostructures \cite{23}. May be the important concept in the
framework of PDM concerns the problem of choosing some fixed ordering
prescription in the kinetic part of the Hamiltonian, coming from
noncommutativity between the mass-term $m(x)$ and momentum $\widehat{P}$ operator.

This problem is a long standing one in quantum mechanics and up to now some
uncertainties are concerned about the form of the kinetic energy term
$\widehat{T}$ in the Hamiltonian. To cope with this difficulty, it was
stressed by von Roos \cite{24} that the correct kinetic part of the
Hamiltonian can be written as%
\begin{equation}
h_{\text{vR}}=\frac{1}{4}\left(  M^{a^{\prime}}(x)\widehat{P}M^{b^{\prime}%
}(x)\widehat{P}M^{c^{\prime}}(x)+M^{c^{\prime}}(x)\widehat{P}M^{b^{\prime}%
}(x)\widehat{P}M^{a^{\prime}}(x)\right)  +V(x),\tag{1.1}\label{1.1}%
\end{equation}
which has an advantage to keep $h_{\text{vR}}$ hermitian. Here $V(x)$ is a
potential and $\widehat{P}\left(  \equiv-i\hbar\frac{d}{dx}\right)  $ is the
conventional momentum operator. $M(x)=m_{0}m(x)$ is the position-dependent
mass function ($m(x)$ being dimensionless and $m_{0}$ is a constant mass), and
$a^{\prime}$, $b^{\prime}$ and $c^{\prime}$ are arbitrary parameters
satisfying the condition: $a^{\prime}+b^{\prime}+c^{\prime}=-1$.

In the other context, a set of coherent states (CS) may be the most important
set of basis vectors knows in mathematical physics and which are not
necessarily orthonormals. Originally, they were first introduced for the
harmonic oscillator by Schr\"{o}dinger in 1926 \cite{25}, and latter by
Glauber \cite{26} who showed that these states describe the electromagnetic
correlation functions in the quantum optics. These states are also constructed
for a family of isospectral oscillator Hamiltonians in \cite{27}. The
construction of CS is approached by three different ways and lead to the
equivalent state vector for harmonic oscillator (see, e.g., \cite{28} and
references therein). For instance, we can recall their definition: (i) as
eigenstates of an annihilation operator (Barut-Girardello's approach), (ii) as
a displaced version of the ground-state wave-function (Klauder-Perelomov's
approach) and (iii) as minimum uncertainty states (Schr\"{o}dinger's
approach). Very recently, CS have also been constructed for a wide family of
Hamiltonians endowed with PDM \cite{29,30,31,32}.

In this paper, our primary concern is to extend the procedure worked out in
\cite{33} to PDM endowed with an effective potential and construct their PDM
CS. We want to tackle the problem of ordering ambiguity in a new way and
differently than \cite{31}, and prove that there is a special ordering in PDM
framework due essentially to the existence of the unitary transformation. Once
the suitable factorization operators, $\mathcal{Q}_{\alpha}$ and
$\mathcal{Q}_{\alpha}^{\dag}$, have been chosen to work as ladders, we prove
that selecting the appropriate SUSY-parameter values $\alpha$ can be
interpreted as a consequence of both \textit{exact SUSY and accidental
symmetries}. We show that the last symmetry is implemented under the $%
\mathbb{Z}
_{2}$-transformation, through an operator $\widehat{\mathcal{Z}}$, if and only
if $\alpha=1/2$ is considered as \textit{the unique fixed point of the }$%
\mathbb{Z}
_{2}$\textit{ action}. The related PDM CS are then constructed as a displaced
version of the ground-state wave-function and minimize the generalized
position-momentum uncertainty relation (GUR). The properties of GUR lead us to
remark two important observations: (i) if $\alpha=1/2$ then the deduced PDM CS
saturate completely GUR and (ii) PDM CS are minimized when $\alpha\neq1/2.$ We
observe that the minimization allow us to fix the sign of the superpotential
$\mathcal{W}_{\alpha}(x)$ appearing in both ladder operators.

The paper is organized as follows. We start by introducing in section 2 the
family of PDM isospectral Hamiltonians, their eigenstates and a pair of
associated ladder operators through the use of unitary transformation. In
section 3 we show that there is a special ordering, that may be checked by a
one SUSY parameter, to select the kinetic part of the Hamiltonian. The related
PDM CS are constructed in section 4. They are shown to preserve the structure
of Klauder-Perelomov's states, and saturate and minimize GUR under some
special restrictions. Our concluding remarks are summarized in the last section.

\section{Effective mass ladder operators and unitary transformation}

Here we follow Morrow and Brownstein \cite{34} who have shown from (\ref{1.1})
the constraint $a^{\prime}=c^{\prime}$, by comparing exact solutions of some
models with experimental results. The restricted Hamiltonian (\ref{1.1}), with
a new change in parameters $a^{\prime}=a$ and $b^{\prime}=2b$, can be now
written as (see, \cite{31})%
\begin{equation}
h_{a}=\frac{1}{2}\left(  m^{a}(x)\widehat{P}m^{2b}(x)\widehat{P}%
m^{a}(x)\right)  +V(x), \tag{2.1}\label{2.1}%
\end{equation}
where we used the atomic units $\hbar=m_{0}=1$ and $2a+2b=-1.$ Let us
factorize (\ref{2.1}) in terms of the operators%
\begin{align}
q_{a}  &  =\frac{1}{\sqrt{2}}\left(  m^{a}(x)\frac{d}{dx}m^{b}(x)+w(x)\right)
,\tag{2.2a}\label{2.2a}\\
q_{a}^{\dag}  &  =\frac{1}{\sqrt{2}}\left(  -m^{b}(x)\frac{d}{dx}%
m^{a}(x)+w(x)\right)  , \tag{2.2b}\label{2.2b}%
\end{align}
where $w(x)$ is an arbitrary real-function. The Hamiltonians $q_{a}^{\dag
}q_{a}$ and $q_{a}q_{a}^{\dag}$ are SUSY partners. It is well known that
$q_{a}$ and $q_{a}^{\dag}$ are not unique when the only requirement for
$h_{a}$ is its hermiticity, then one can construct a family of strictly
isospectral Hamiltonians through a new operators \cite{33}%
\begin{align}
Q_{a}  &  =\frac{1}{\sqrt{2}}\left(  m^{a}(x)\frac{d}{dx}m^{b}(x)+W(x)\right)
,\tag{2.3a}\label{2.3a}\\
Q_{a}^{\dag}  &  =\frac{1}{\sqrt{2}}\left(  -m^{b}(x)\frac{d}{dx}%
m^{a}(x)+W(x)\right)  , \tag{2.3b}\label{2.3b}%
\end{align}
such that%
\begin{equation}
Q_{a}Q_{a}^{\dag}=q_{a}q_{a}^{\dag}. \tag{2.4}\label{2.4}%
\end{equation}

On setting $m(x)=U^{-2}(x)$, where $U(x)$ is some positive-definite function
and substituting (2.2) and (2.3) into (\ref{2.4}), one gets Riccati equation
which can be solved to give:%
\begin{equation}
W(x)=w(x)+\phi_{\lambda}(x),\qquad\text{with\qquad}\phi_{\lambda}%
(x)=\frac{U(x)}{\lambda+%
{\displaystyle\int\nolimits_{{}}^{x}}
\xi_{0}^{2}(\eta)d\eta}\xi_{0}^{2}(x), \tag{2.5}\label{2.5}%
\end{equation}
where $\lambda$ is a constant of integration and $\xi_{0}(x)$ is the
normalized ground-state eigenfunction of the Hamiltonian $q_{a}^{\dag}q_{a}$,
given by%
\[
\xi_{0}(x)\sim U^{-1-2a}(x)\exp\left\{  -%
{\displaystyle\int\nolimits^{\mu(x)}}
w(\eta)d\mu(\eta)\right\}  ,
\]
up to normalization constant and for convenience we introduce the auxiliary
mass function $\mu(x)=%
{\textstyle\int\nolimits^{x}}
d\eta/U(\eta)$.

One can construct the eigenfunctions $\left\vert \Xi_{n}\right\rangle $ of the
strictly isospectral family of $H_{a}=Q_{a}^{\dag}Q_{a}+\epsilon$, ($\epsilon$
corresponds to the ground-state energy, i.e., $\epsilon=E_{0}$), which is
intertwined with $h_{a}=q_{a}^{\dag}q_{a}+\epsilon$ by means of the
intertwining operator $\mathcal{B}_{a}=Q_{a}^{\dag}q_{a}$ \cite{13,33}
\begin{equation}
H_{a}\mathcal{B}_{a}=\mathcal{B}_{a}h_{a}\quad\Longrightarrow\quad\left(
Q_{a}^{\dag}Q_{a}\right)  Q_{a}^{\dag}q_{a}=Q_{a}^{\dag}q_{a}\left(
q_{a}^{\dag}q_{a}\right)  . \tag{2.6}\label{2.6}%
\end{equation}

It turns out that the normalized eigenfunctions $\left\vert \Xi_{n}%
\right\rangle $ of $H_{a}$ can be written in terms of the eigenfunctions
$\left\vert \xi_{n}\right\rangle $ as%
\begin{equation}
\left\vert \Xi_{n}\right\rangle =\frac{1}{E_{n}-\epsilon}Q_{a}^{\dag}%
q_{a}\left\vert \xi_{n}\right\rangle ,\qquad(n=1,2,3,\cdots) \tag{2.7}%
\label{2.7}%
\end{equation}

It is worth noting from (\ref{2.7}) that $\epsilon=E_{0}\notin$ Spect$\left(
H_{a}\right)  $. From (\ref{2.7}), it is easy to determine the normalized
eigenfunctions $\Xi_{n}(x)\left(  \equiv\left\langle x|\Xi_{n}\right\rangle
\right)  $ in the coordinate representation and are given by%
\begin{align}
\Xi_{0}(x)  &  \sim\frac{\xi_{0}(x)}{\lambda+%
{\displaystyle\int\nolimits_{{}}^{x}}
\xi_{0}^{2}(\eta)d\eta},\tag{2.8}\label{2.8}\\
\Xi_{n}(x)  &  =\left(  1+\frac{1}{2(E_{n}-\epsilon)}\phi_{\lambda}%
(x)\ q_{a}\right)  \xi_{n}(x). \tag{2.9}\label{2.9}%
\end{align}

Thus the operators $Q_{a}$ and $q_{a}$ connect the states $\left\vert \xi
_{n}\right\rangle $ and $\left\vert \Xi_{n}\right\rangle $ and can not be
considered as creation and annihilation operators. To cope with this
difficulty, we are interested now in identifying these set of ladder operators
by introducing the unitary transformation, $\widehat{\mathfrak{U}}$, following
Kumar and Khare \cite{33}%
\begin{equation}
Q_{a}=q_{a}\widehat{\mathfrak{U}}^{\dag},\qquad\text{and\qquad}Q_{a}^{\dag
}=\widehat{\mathfrak{U}}q_{a}^{\dag}, \tag{2.10}\label{2.10}%
\end{equation}
where (\ref{2.4}) implies that $\widehat{\mathfrak{U}}^{\dag}\widehat
{\mathfrak{U}}\equiv\widehat{\mathfrak{U}}\widehat{\mathfrak{U}}^{\dag
}=\openone$. Once more, let us consider the factorization%
\begin{equation}
\mathcal{H}_{a}=\mathcal{Q}_{a}^{\dag}\mathcal{Q}_{a}+\epsilon, \tag{2.11}%
\label{2.11}%
\end{equation}
by defining the new ladder operators $\mathcal{Q}_{a}$ and $\mathcal{Q}%
_{a}^{\dag}$ in terms of $\widehat{\mathfrak{U}}$ and $\widehat{\mathfrak{U}%
}^{\dag}$ as%
\begin{equation}
\mathcal{Q}_{a}=\widehat{\mathfrak{U}}q_{a}\widehat{\mathfrak{U}}^{\dag}%
\equiv\widehat{\mathfrak{U}}Q_{a},\qquad\text{and\qquad}\mathcal{Q}_{a}^{\dag
}=\widehat{\mathfrak{U}}q_{a}^{\dag}\widehat{\mathfrak{U}}^{\dag}\equiv
Q_{a}^{\dag}\widehat{\mathfrak{U}}^{\dag}, \tag{2.12}\label{2.12}%
\end{equation}
so that $Q_{a}^{\dag}Q_{a}=\mathcal{Q}_{a}^{\dag}\mathcal{Q}_{a}%
\colon=\widehat{\mathfrak{U}}\left(  q_{a}^{\dag}q_{a}\right)  \widehat
{\mathfrak{U}}^{\dag}$ and leads to the relation $\mathcal{H}_{a}%
=\widehat{\mathfrak{U}}h_{a}\widehat{\mathfrak{U}}^{\dag}$. A consequence of
isospectrality between $h_{a}$ and $H_{a}$ implies that $E_{n}\equiv
\left\langle \Xi_{n}\left\vert H_{a}\right\vert \Xi_{n}\right\rangle
=\left\langle \xi_{n}\left\vert h_{a}\right\vert \xi_{n}\right\rangle $.

Using (\ref{2.12}) we get a relation connecting the basis $\left\{  \left\vert
\xi_{n}\right\rangle \right\}  _{n\in%
\mathbb{N}
}$ and $\left\{  \left\vert \Xi_{n}\right\rangle \right\}  _{n\in%
\mathbb{N}
}$ as follows%
\begin{equation}
\left\vert \xi_{n}\right\rangle =\widehat{\mathfrak{U}}^{\dag}\left\vert
\Xi_{n}\right\rangle ,\qquad\text{and\qquad}\left\vert \Xi_{n}\right\rangle
=\widehat{\mathfrak{U}}\left\vert \xi_{n}\right\rangle . \tag{2.13}%
\label{2.13}%
\end{equation}

Moreover, the relevance of relationships (\ref{2.10}), (\ref{2.12}) and
(\ref{2.13}) is clear by noting that
\begin{align}
\left\langle \xi_{n}\left\vert q_{a}^{\dag}q_{a}\right\vert \xi_{n}%
\right\rangle  &  =\left\langle \xi_{n}\left\vert \widehat{\mathfrak{U}}%
^{\dag}\left(  \widehat{\mathfrak{U}}q_{a}^{\dag}\right)  \left(
q_{a}\widehat{\mathfrak{U}}^{\dag}\right)  \widehat{\mathfrak{U}}\right\vert
\xi_{n}\right\rangle \nonumber\\
&  =\left\langle \widehat{\mathfrak{U}}\xi_{n}\left\vert Q_{a}^{\dag}%
Q_{a}\right\vert \widehat{\mathfrak{U}}\xi_{n}\right\rangle \nonumber\\
&  =\left\langle Q_{a}\widehat{\mathfrak{U}}\xi_{n}|Q_{a}\widehat
{\mathfrak{U}}\xi_{n}\right\rangle \equiv E_{n}-\epsilon,\tag{2.14}%
\label{2.14}%
\end{align}
which leads to define a new state $\left\vert \Upsilon_{n}\right\rangle
\propto Q_{a}\left(  \widehat{\mathfrak{U}}\left\vert \xi_{n}\right\rangle
\right)  =Q_{a}\left\vert \Xi_{n}\right\rangle $, such that $\left\langle
\Upsilon_{n}|\Upsilon_{n}\right\rangle =\left\Vert \left\vert \Upsilon
_{n}\right\rangle \right\Vert ^{2}=E_{n}-\epsilon<\infty$. Then, the new set
of states%
\begin{equation}
S_{n}=\left\{  \left\vert \Upsilon_{n}\right\rangle =\frac{1}{\sqrt
{E_{n}-\epsilon}}Q_{a}\left(  \widehat{\mathfrak{U}}\left\vert \xi
_{n}\right\rangle \right)  =\frac{1}{\sqrt{E_{n}-\epsilon}}Q_{a}\left\vert
\Xi_{n}\right\rangle
\bigg{|}%
E_{n}\neq\epsilon\right\}  ,\tag{2.15}\label{2.15}%
\end{equation}
consists of normalized eigenfunctions of $H_{a}\equiv\mathcal{H}%
_{a}=\mathcal{Q}_{a}^{\dag}\mathcal{Q}_{a}+\epsilon$ belonging to the
eigenvalues $E_{n}=$ Spect$(H_{a})$. Now, following \cite{31}, let $\left\vert
\Upsilon_{n,\epsilon}\right\rangle $ be a function which is orthogonal to the
set $\left\{  \left\vert \Upsilon_{n}\right\rangle \right\}  _{n\in%
\mathbb{N}
}$; i.e., $\left\langle \Upsilon_{n}|\Upsilon_{n,\epsilon}\right\rangle
\equiv\left\langle Q_{a}\Xi_{n}|\Upsilon_{n,\epsilon}\right\rangle
=\left\langle \Xi_{n}|Q_{a}^{\dag}\Upsilon_{n,\epsilon}\right\rangle =0$. Then
using (\ref{2.10}), it is easy to convince ourselves that $Q_{a}^{\dag
}\left\vert \Upsilon_{n,\epsilon}\right\rangle \equiv\widehat{\mathfrak{U}%
}\left(  q_{a}^{\dag}\left\vert \Upsilon_{n,\epsilon}\right\rangle \right)
=0$, since the state $\left\vert \Xi_{n}\right\rangle \neq0$ and the related
solution reads, in the coordinate representation, as%
\begin{equation}
\Upsilon_{n,\epsilon}(x)\sim\frac{1}{m^{a}(x)}\exp\left\{
{\displaystyle\int\nolimits^{\mu(x)}}
w(\eta)d\mu\left(  \eta\right)  \right\}  .\tag{2.16}\label{2.16}%
\end{equation}

If $\left\langle \Upsilon_{n,\epsilon}|\Upsilon_{n,\epsilon}\right\rangle
<\infty$, then Spect$(H_{a})=$ Spect$(h_{a})\cup\left\{  \epsilon\right\}  $
which leads to impose that $m(x)$ \textit{has no zeros }and $a<0$ (i.e.,
$b>-1/2$). This problem is closely related to the ordering problem and we will
face to this question in the next section.

In summary, $h_{a}$ and $H_{a}$ are isospectral operators and are related by
the unitary transformation $\widehat{\mathfrak{U}}$. We would like also to
emphasize that $\mathcal{Q}_{a}$ and $\mathcal{Q}_{a}^{\dag}$, as they are
defined in (\ref{2.12}), are nothing but the correct set of creation and
annihilation operators for our isospectral Hamiltonians.

\section{SUSY parameter $\alpha$ and accidental symmetry $%
\mathbb{Z}
_{2}$}

In this section we take full advantage of ladder operators derived under
unitary transformation in the previous section. To this end, inserting
(\ref{2.2a}) into (\ref{2.12}), the operator $\mathcal{Q}_{a}$ can be
expressed as%
\begin{equation}
\mathcal{Q}_{a}=\widehat{\mathfrak{U}}q_{a}\widehat{\mathfrak{U}}^{\dag}%
\equiv\frac{1}{\sqrt{2}}\left(  \widehat{\mathfrak{U}}m^{a}(x)\frac{d}%
{dx}m^{b}(x)\widehat{\mathfrak{U}}^{\dag}+\widehat{\mathfrak{U}}%
w(x)\widehat{\mathfrak{U}}^{\dag}\right)  .\tag{3.1}\label{3.1}%
\end{equation}

By substituting$\ \widehat{\mathfrak{U}}^{\dag}\widehat{\mathfrak{U}%
}=\openone$ twice into the first term of the right-hand side of (\ref{3.1}),
taking into account that $b=-(1/2+a)$, the operators $\mathcal{Q}_{a}$ and
$\mathcal{Q}_{a}^{\dag}$ can be rewritten explicitly in the form:%
\begin{align}
\mathcal{Q}_{a} &  \longrightarrow\mathcal{Q}_{\alpha}=\frac{1}{\sqrt{2}%
}\left(  f^{\alpha}(x)\frac{d}{dx}g(x)f^{-\alpha}(x)+\mathcal{W}(x)\right)
,\tag{3.2}\label{3.2}\\
\mathcal{Q}_{a}^{\dag} &  \longrightarrow\mathcal{Q}_{\alpha}^{\dag}=\frac
{1}{\sqrt{2}}\left(  -g(x)f^{-\alpha}(x)\frac{d}{dx}f^{\alpha}(x)+\mathcal{W}%
(x)\right)  ,\tag{3.3}\label{3.3}%
\end{align}
where $f(x)$, $g(x)$ and the superpotential $\mathcal{W}(x)$ are real
functions defined following%
\begin{equation}
f^{\alpha}(x)=\widehat{\mathfrak{U}}f^{a}(x)\widehat{\mathfrak{U}}^{\dag
},\qquad g(x)=\widehat{\mathfrak{U}}m^{-1/2}(x)\widehat{\mathfrak{U}}^{\dag
},\qquad\text{and\qquad}\mathcal{W}(x)=\widehat{\mathfrak{U}}w(x)\widehat
{\mathfrak{U}}^{\dag},\tag{3.4}\label{3.4}%
\end{equation}
where we assume that the new parameter $\alpha$ only labels a particular
ordering in the kinetic part of the Hamiltonian. Then from (\ref{2.11}),
$\mathcal{H}_{a}\rightarrow\mathcal{H}_{\alpha}=\mathcal{Q}_{\alpha}^{\dag
}\mathcal{Q}_{\alpha}+\epsilon$, we have%
\begin{equation}
\mathcal{H}_{\alpha}=-\frac{1}{2}\frac{d}{dx}g^{2}(x)\frac{d}{dx}%
+\mathcal{V}_{\text{eff}}^{(\alpha)}(x)+\epsilon,\tag{3.5}\label{3.5}%
\end{equation}
where $\mathcal{V}_{\text{eff}}^{(\alpha)}(x)$ is known as the effective
potential. In order to relate (\ref{3.5}) to PDM problem, we look for the
functions $f(x)$ and $g(x)$ that satisfy the condition $f(x)=g(x)=U(x)$. Then
(\ref{3.2}) and (\ref{3.3}) become%
\begin{subequations}
\begin{align}
\mathcal{Q}_{\alpha} &  \equiv\frac{1}{\sqrt{2}}\left(  U^{\alpha}(x)\frac
{d}{dx}U^{1-\alpha}(x)+\mathcal{W}_{\alpha}(x)\right)  ,\tag{3.6a}%
\label{3.6a}\\
\mathcal{Q}_{\alpha}^{\dag} &  \equiv\frac{1}{\sqrt{2}}\left(  -U^{1-\alpha
}(x)\frac{d}{dx}U^{\alpha}(x)+\mathcal{W}_{\alpha}(x)\right)  ,\tag{3.6b}%
\label{3.6b}%
\end{align}
where the effective potential, $\mathcal{V}_{\text{eff}}^{(\alpha
)}(x)=\mathcal{V}_{\alpha}(x)+\mathcal{V}_{\text{U}}^{(\alpha)}(x)$, is
defined in terms of the superpotential $\mathcal{W}_{\alpha}(x)$ as%
\end{subequations}
\begin{align}
\mathcal{V}_{\alpha}(x)-\epsilon &  =\frac{1}{2}\mathcal{W}_{\alpha}%
^{2}(x)-\frac{1}{2}U(x)\mathcal{W}_{\alpha}^{\prime}(x)+\frac{1-2\alpha}%
{2}U^{\prime}(x)\mathcal{W}_{\alpha}(x),\tag{3.7a}\label{3.7a}\\
\mathcal{V}_{\text{U}}^{(\alpha)}(x) &  =\frac{\alpha(\alpha-1)}{2}U^{\prime
2}(x)+\frac{\alpha-1}{2}U^{\prime\prime}(x)U(x).\tag{3.7b}\label{3.7b}%
\end{align}

However, we have observed that the appropriate mapping: $\alpha\mapsto
\alpha_{n}=n\alpha$, for $n\in%
\mathbb{N}
-\left\{  0\right\}  $, has a remarkable property; it keeps the kinetic
operator in (\ref{3.5}) \textit{unchanged} and we will see hereafter why these
maps were chosen. We shall take full advantage of this last mapping, since the
problem we have to deal with concerns the choice of the appropriate profile
for mass functions $U(x),$ (i.e., $m(x)$), appearing in (3.6) and represented
by the terms $\zeta^{(+)}=U^{\alpha_{n}}(x)$ and $\zeta^{(-)}=U^{1-\alpha_{n}%
}(x)$. Taking into account that $U(x)=m^{-1/2}(x)$, we are faced the situation
in which two distinguishable profiles for $U(x)$ must be considered:

\begin{description}
\item[(P1)] $\zeta^{(\pm)}(x)$\textit{ admit singularities (i.e., }%
$m(x)$\textit{ has zeros)}. This implies that both exponents $(1-n\alpha
,n\alpha)<0$, with $\alpha$ is confined to the domain: dom$(\alpha)=\left(
-\infty,0\right]  \cup\left[  \frac{1}{n},+\infty\right)  $. Then it is easy
to realize that this case must be omitted in order to avoid a possible
divergence of $\mathcal{Q}_{\alpha}$ (i.e., $\mathcal{H}_{\alpha}$), when
$\alpha=\pm\infty$.

\item[(P2)] $\zeta^{(\pm)}(x)$\textit{ admit zeros (i.e., }$m(x)$\textit{ has
singularities)}. This implies that $(1-n\alpha,n\alpha)>0$ and a similar
reasoning shows that $0\leq\alpha\leq\frac{1}{n}\leq1$, for $n=1,2,3,\cdots$.
\end{description}

The case (\textbf{P2}) is which we are looking for. Then it is evident that
the functions $U(x)$ (resp. $m(x)$) are chosen in such a way that they accept
\textit{zeros} (resp. \textit{singularities}). This fact suggest that the
ordering in the kinetic operator depends closely on the choice of SUSY
parameter $\alpha$ which belongs to the discrete and bounded set:%
\begin{equation}
\mathfrak{S}_{\alpha_{n}}=\left\{  \forall n\in%
\mathbb{N}
-\left\{  0\right\}
\bigg{|}%
\text{ }\alpha_{\infty}=0,\quad\text{and\quad}\alpha_{n}=\frac{1}{n}\in%
\mathbb{Q}
^{+}\right\}  \equiv\left\{  1,\frac{1}{2},\frac{1}{3},\ldots,\frac{1}%
{n},\ldots,0\right\}  , \tag{3.8}\label{3.8}%
\end{equation}
which is well-known that it is closed in $%
\mathbb{R}
$, and hence compact. For instances, among the SUSY parameter $\alpha_{n}$ of
the set (\ref{3.8}), some of frequently used form of the kinetic operators are
reported below in the table. In the ongoing analysis, we will suppress the
$n$-index of all the parameters $\alpha$.%

\begin{table}[h] \centering
\caption{Some of mostly used forms of the kinetic operators for describing PDM system. Among those ambiguity parameter choices, we may quote those of: (i) Zhu and Kroemer (ZK) and (ii) BenDaniel and Duke (BDD) both used in describing the motion of electrons in compositionally graded crystals, and (iii) Bagchi, Banerjee, Quesne and Tkachuk (BBQT).}%
\begin{tabular}
[c]{llllll}\hline
\multicolumn{2}{l}{Parameters} & \multicolumn{2}{c}{Operators} &  &
\\\cline{3-4}%
$(\alpha,n)$ & $(a,b)$ & $\mathcal{Q}_{\alpha}$ & $\widehat{T}_{\alpha}$ &
(Type) & Refs.\\\hline
$(0,\infty)$ & $\left(  -\frac{1}{2},0\right)  $ & $\frac{1}{\sqrt{2}}\left(
\frac{d}{dx}U(x)+\mathcal{W}_{0}(x)\right)  $ & $\frac{1}{2}U(x)\widehat
{P}^{2}U(x)$ & (ZK) & \cite{35}\\
$(1,1)$ & $\left(  0,-\frac{1}{2}\right)  $ & $\frac{1}{\sqrt{2}}\left(
U(x)\frac{d}{dx}+\mathcal{W}_{1}(x)\right)  $ & $\frac{1}{2}\widehat{P}%
U^{2}(x)\widehat{P}$ & (BDD) & \cite{36}\\
$\left(  \frac{1}{2},2\right)  $ & $\left(  -\frac{1}{4},-\frac{1}{4}\right)
$ & $\frac{1}{\sqrt{2}}\left(  \sqrt{U(x)}\frac{d}{dx}\sqrt{U(x)}%
+\mathcal{W}_{1/2}(x)\right)  $ & $\frac{1}{2}\sqrt{U(x)}\widehat
{P}U(x)\widehat{P}\sqrt{U(x)}$ & (BBQT) & \cite{11}\\\hline
\end{tabular}
\label{table1}%
\end{table}%

Using the identity (\ref{2.11}) and the representations (\ref{3.2}) and
(\ref{3.3}), one can easily find that%
\begin{align}
\left[  \mathcal{Q}_{\alpha},\mathcal{Q}_{\alpha}^{\dag}\right]   &
=U(x)\mathcal{W}_{\alpha}^{\prime}(x)+\frac{1-2\alpha}{2}U(x)U^{\prime\prime
}(x),\tag{3.9a}\label{3.9a}\\
\left[  \mathcal{H}_{\alpha},\mathcal{Q}_{\alpha}\right]   &  =-\left[
\mathcal{Q}_{\alpha},\mathcal{Q}_{\alpha}^{\dag}\right]  \mathcal{Q}_{\alpha
},\tag{3.9b}\label{3.9b}\\
\left[  \mathcal{H}_{\alpha},\mathcal{Q}_{\alpha}^{\dag}\right]   &
=\mathcal{Q}_{\alpha}^{\dag}\left[  \mathcal{Q}_{\alpha},\mathcal{Q}_{\alpha
}^{\dag}\right]  , \tag{3.9c}\label{3.9c}%
\end{align}
which require that the algebra of three operators $\mathcal{Q}_{\alpha
},\mathcal{Q}_{\alpha}^{\dag}$ and $\mathcal{H}_{\alpha}$ is nothing but the
generally deformed oscillator algebra (GDOA) \cite{29}.

Although our approach holds for any $\alpha$ of (\ref{3.8}), we shall focus
our attention on the effect changing the SUSY parameter $\alpha$ to $1-\alpha$
via the identification: $\alpha\mapsto\overline{\alpha}=1-\alpha$, with
$0\leq\overline{\alpha}\leq1$. Then (3.6) become%
\begin{subequations}
\begin{align}
\mathcal{Q}_{\alpha}  &  \rightarrow\overline{\mathcal{Q}}_{\alpha}%
\equiv\mathcal{Q}_{1-\alpha}=\frac{1}{\sqrt{2}}\left(  U^{1-\alpha}(x)\frac
{d}{dx}U^{\alpha}(x)+\mathcal{W}_{1-\alpha}(x)\right)  , \tag{3.10a}%
\label{3.10a}\\
\mathcal{Q}_{\alpha}^{\dag}  &  \rightarrow\overline{\mathcal{Q}}_{\alpha
}^{\dag}\equiv\mathcal{Q}_{1-\alpha}^{\dag}=\frac{1}{\sqrt{2}}\left(
-U^{\alpha}(x)\frac{d}{dx}U^{1-\alpha}(x)+\mathcal{W}_{1-\alpha}(x)\right)  ,
\tag{3.10b}\label{3.10b}%
\end{align}
where the corresponding effective potential%
\end{subequations}
\begin{align}
\overline{\mathcal{V}}_{\alpha}(x)-\epsilon &  =\frac{1}{2}\mathcal{W}%
_{1-\alpha}^{2}(x)-\frac{1}{2}U(x)\mathcal{W}_{1-\alpha}^{\prime}%
(x)-\frac{1-2\alpha}{2}U^{\prime}(x)\mathcal{W}_{1-\alpha}(x), \tag{3.11a}%
\label{3.11a}\\
\overline{\mathcal{V}}_{\text{U}}^{(\alpha)}(x)  &  =\frac{\alpha(\alpha
-1)}{2}U^{\prime2}(x)-\frac{\alpha}{2}U^{\prime\prime}(x)U(x), \tag{3.11b}%
\label{3.11b}%
\end{align}
and the associated GDOA is the same as (3.9) except that they differ in the
sign of the term $1-2\alpha$.

There is another side to this identification. Indeed a simple inspection leads
us to make a crucial observation that a such changing in the SUSY parameter
$\alpha$ is nothing but an accidental symmetry under the $%
\mathbb{Z}
_{2}$ action, \textit{if and only if }$\alpha=1/2$\textit{ is the unique fixed
point of this transformation}. For example, the emergence of this kind of
striking symmetry has already been discussed by Fiset and Hussin \cite{37}.
They find that this symmetry has an interesting and particular effects on the
coherent states built out of SUSY potential functions. Here by an accidental
symmetry, we mean that this symmetry in not predicted neither by factorization
method, nor the SUSY approach. As we can see, selecting an appropriate SUSY
parameter $\alpha$ can also be interpreted as a consequence of the striking
and accidental symmetry under the $%
\mathbb{Z}
_{2}$ action, where only $\overline{\alpha}=0,1/2$, and $1$ belong to the set
(\ref{3.8}). It is obvious that the most suitable ordering parameter $\alpha$
is given by the simplest case $\alpha=1/2$, (i.e., $a=b=-1/4$, see Table 1),
and corresponding to the well-known kinetic operator, $\widehat{T}_{1/2}%
=\frac{1}{2}\left(  \sqrt{U(x)}\frac{d}{dx}\sqrt{U(x)}\right)  ^{2}$, which
has deserved special attention in research papers over the years (see, for
example \cite{11} and references therein.)

To be more precise about this transformation, let us introduce an operator
$\widehat{\mathcal{Z}}$ that implements the $%
\mathbb{Z}
_{2}$-transformation on the annihilation operator $\mathcal{Q}_{\alpha}$.
Thus, acting on $\mathcal{Q}_{\alpha}$, we must require that the implement
operator $\widehat{\mathcal{Z}}$ satisfies%
\begin{equation}
\widehat{\mathcal{Z}}\mathcal{Q}_{\alpha}\widehat{\mathcal{Z}}^{-1}%
=\mathcal{Q}_{1-\alpha}\quad\Longrightarrow\quad\widehat{\mathcal{Z}%
}\mathcal{Q}_{\alpha}=\mathcal{Q}_{1-\alpha}\widehat{\mathcal{Z}},
\tag{3.13}\label{3.13}%
\end{equation}
with $\widehat{\mathcal{Z}}^{-1}\neq\widehat{\mathcal{Z}}^{\dag}$, since we
assume that they are no other operators, apart from $\mathcal{Q}_{\alpha}$,
which are affected by the $%
\mathbb{Z}
_{2}$ action. Note that we have also demand the invariance for the fixed point
$\alpha=1/2$, i.e., the commutator $\left[  \widehat{\mathcal{Z}}%
,\mathcal{Q}_{1/2}\right]  =0$, which means that the operators $\widehat
{\mathcal{Z}}$ and $\mathcal{Q}_{1/2}$ share the same eigenfunctions. Then,
along with (\ref{3.13}), it is obvious that the implement operator
$\widehat{\mathcal{Z}}$ is identified to $\mathcal{Q}_{1/2}$ for $\alpha=1/2$.
However it acts as an intertwiner transformation for the ladder operators
$\mathcal{Q}_{\alpha}$ and $\mathcal{Q}_{1-\alpha}$ for $\alpha\neq1/2$.
Moreover, acting (\ref{3.13}) on the right-hand side by a function $\phi
_{n}(x)$, we get%
\begin{equation}
\Psi_{n}(x)\equiv\mathcal{Q}_{1-\alpha}\psi_{n}(x)=\widehat{\mathcal{Z}%
}\left(  \mathcal{Q}_{\alpha}\varphi_{n}(x)\right)  \equiv\widehat
{\mathcal{Z}}\Phi_{n}(x), \tag{3.14}\label{3.14}%
\end{equation}
where $\psi_{n}(x)\equiv\widehat{\mathcal{Z}}\varphi_{n}(x)$. Clearly, we look
for $\widehat{\mathcal{Z}}$ as an \textit{intertwiner operator}. For the sake
of completeness, we now seek the intertwining operator in the form of a
first-order linear-differential operator%
\begin{equation}
\widehat{\mathcal{Z}}=\frac{1}{\sqrt{2}}\left(  F(x)\frac{d}{dx}F\left(
x\right)  +G(x)\right)  , \tag{3.15}\label{3.15}%
\end{equation}
where $F(x)$ and $G(x)$ are two real functions to be determined, such that
$\widehat{\mathcal{Z}}$ fulfills (\ref{3.13}). For lack of space, the
implement operator is evaluated and the result is%
\begin{equation}
\widehat{\mathcal{Z}}\equiv\mathcal{T}_{\alpha}(x)\widehat{\nabla}%
_{x}^{(\alpha)}\mathcal{T}_{\alpha}(x), \tag{3.16}\label{3.16}%
\end{equation}
with%
\begin{align}
\mathcal{T}_{\alpha}(x)  &  =U^{1-\alpha}(x)\exp\left\{  \int_{{}}^{\mu
(x)}\frac{\mathcal{W}_{\alpha}(\eta)-\mathcal{W}_{1-\alpha}(\eta)}{2}d\mu
(\eta)\right\}  ,\qquad\text{and}\tag{3.17a}\label{3.17a}\\
\widehat{\nabla}_{x}^{(\alpha)}  &  =\frac{1}{\sqrt{2}}\left(  \frac{d}%
{dx}+\frac{1}{2}\frac{\mathcal{W}_{\alpha}(x)+\mathcal{W}_{1-\alpha}(x)}%
{U(x)}\right)  . \tag{3.17b}\label{3.17b}%
\end{align}

It is worth noting that if $\alpha=1/2$ , then (\ref{3.17a}) is reduced to
$\mathcal{T}_{1/2}(x)=U^{1/2}(x)\equiv m^{-1/4}(x)$ and the implement operator
$\widehat{\mathcal{Z}}$ becomes the standard annihilation operator
$\mathcal{Q}_{1/2}$, (see Table 1), i.e.,%
\begin{equation}
\widehat{\mathcal{Z}}=\mathcal{Q}_{1/2}=\frac{1}{\sqrt{2}}\left(  \sqrt
{U(x)}\frac{d}{dx}\sqrt{U(x)}+\mathcal{W}_{1/2}(x)\right)  . \tag{3.18}%
\label{3.18}%
\end{equation}

\section{PDM CS and generalized position-momentum uncertainty relation}

Now the main question is: what kind of position-dependent mass coherent states
(PDM CS) are expected in both symmetries? This section is devoted to construct
a set of PDM CS for isospectral Hamiltonians $\mathcal{H}_{\alpha}$. As
mentioned in the introduction, there are three equivalent definitions of CS
and one of the candidates (Klauder-Perelomov's approach) looks on CS as an
orbit of the ground-state $\left\vert 0\right\rangle $, under the
Weyl-Heisenberg displacement operator $\mathcal{D}(z)$ expressed in terms of
ladder operators.

We have seen that $\mathcal{Q}_{\alpha}$ and $\mathcal{Q}_{\alpha}^{\dag}$ are
such operators and can be used to derive their associated PDM CS. Then the
related PDM CS, $\left\vert z;\alpha\right\rangle $, for which we are looking
for must verify the eigenvalue equation $\mathcal{Q}_{\alpha}\left\vert
z;\alpha\right\rangle =z\left\vert z;\alpha\right\rangle $ and annihilates the
ground-state $\left\vert \Xi_{0};\alpha\right\rangle $, i.e., $\mathcal{Q}%
_{\alpha}\left\vert \Xi_{0};\alpha\right\rangle =0$.

Using (3.6) the ground-state $\left\vert \Xi_{0};\alpha\right\rangle $ can be
calculated straightforwardly by integration, and we get%
\begin{equation}
\left\vert \Xi_{0};\alpha\right\rangle \sim U^{\alpha-1}(x)\exp\left\{  -%
{\displaystyle\int\nolimits_{{}}^{\mu(x)}}
\mathcal{W}_{\alpha}(\eta)d\mu(\eta)\right\}  , \tag{4.1}\label{4.1}%
\end{equation}
up to normalization constant. In Klauder-Perelomov's approach, our PDM CS are
expressed through%
\begin{equation}
\left\vert z;\alpha\right\rangle =\mathcal{D}_{\alpha}(z)\left\vert \Xi
_{0};\alpha\right\rangle , \tag{4.2}\label{4.2}%
\end{equation}
and one assumes that the displacement operator, $\mathcal{D}_{\alpha}(z)$,
takes the form \cite{29}%
\begin{equation}
\mathcal{D}_{\alpha}(z)=\exp\left\{  i\mathcal{S}_{\alpha}(z)\right\}
,\qquad\text{with\qquad}\mathcal{S}_{\alpha}(z)=-iz\mathcal{K}_{\alpha},
\tag{4.3}\label{4.3}%
\end{equation}
where $\mathcal{K}_{\alpha}$ is unknown operator which can be determined using
the unitarity condition of $\mathcal{D}_{\alpha}(z)$, i.e., $\mathcal{D}%
_{\alpha}^{-1}(z)=\mathcal{D}_{\alpha}^{\dag}(z)$. Indeed this restriction
leads us to identity both $\mathcal{S}_{\alpha}(z)$ and $\mathcal{K}_{\alpha}$
as hermitian operators if and only if $z=-z^{\ast}$, i.e., $\Re(z)=0$. Having
introduced the form of $\mathcal{D}_{\alpha}(z)$, we are now able to state the
following proposition \cite{29} in order to determine $\mathcal{K}_{\alpha}$.

\begin{proposition}
Let $z\in i%
\mathbb{R}
$. Then, for every displacement operator $\mathcal{D}_{\alpha}(z)$ as defined
in (\ref{4.3}) and acting on the ladder operators under the general scheme:%
\begin{equation}
\mathcal{D}_{\alpha}^{\dag}(z)\mathcal{Q}_{\alpha}\mathcal{D}_{\alpha
}(z)=\mathcal{Q}_{\alpha}+z\left[  \mathcal{Q}_{\alpha},\mathcal{Q}_{\alpha
}^{\dag}\right]  ,\qquad\text{and\qquad}\mathcal{D}_{\alpha}^{\dag
}(z)\mathcal{Q}_{\alpha}^{\dag}\mathcal{D}_{\alpha}(z)=\mathcal{Q}_{\alpha
}^{\dag}+z^{\ast}\left[  \mathcal{Q}_{\alpha},\mathcal{Q}_{\alpha}^{\dag
}\right]  , \tag{4.4}\label{4.4}%
\end{equation}
must verify%
\begin{equation}
\left[  \mathcal{S}_{\alpha}(z),\mathcal{Q}_{\alpha}\right]  =iz\left[
\mathcal{Q}_{\alpha},\mathcal{Q}_{\alpha}^{\dag}\right]  ,\qquad
\text{and}\qquad\left[  \mathcal{S}_{\alpha}(z),\mathcal{Q}_{\alpha}^{\dag
}\right]  =iz^{\ast}\left[  \mathcal{Q}_{\alpha},\mathcal{Q}_{\alpha}^{\dag
}\right]  , \tag{4.5}\label{4.5}%
\end{equation}
where $z=-z^{\ast}$.

\begin{proof}
Let us define the operator $\mathcal{P}_{\alpha}=\left[  \mathcal{Q}_{\alpha
},\mathcal{D}_{\alpha}(z)\right]  $. Expanding $\exp\left\{  i\mathcal{S}%
_{\alpha}(z)\right\}  $ in the development of Taylor series, we find%
\begin{equation}
\mathcal{P}_{\alpha}=-\sum_{k=0}^{+\infty}\frac{i^{k}}{k!}\left[
\mathcal{S}_{\alpha}^{k}(z),\mathcal{Q}_{\alpha}\right]  , \tag{4.6}%
\label{4.6}%
\end{equation}
and using (\ref{4.5}), a straightforward calculation yields to the recursion
relation satisfying%
\begin{equation}
\left[  \mathcal{S}_{\alpha}^{k}(z),\mathcal{Q}_{\alpha}\right]
=ik\mathcal{S}_{\alpha}^{k-1}(z)\left[  \mathcal{Q}_{\alpha},\mathcal{Q}%
_{\alpha}^{\dag}\right]  , \tag{4.7}\label{4.7}%
\end{equation}
and by inserting (\ref{4.7}) into (\ref{4.6}), we have%
\begin{equation}
\mathcal{P}_{\alpha}=z\mathcal{D}_{\alpha}(z)\left[  \mathcal{Q}_{\alpha
},\mathcal{Q}_{\alpha}^{\dag}\right]  . \tag{4.8}\label{4.8}%
\end{equation}

On the other hand, starting from $\mathcal{P}_{\alpha}=\left[  \mathcal{Q}%
_{\alpha},\mathcal{D}_{\alpha}(z)\right]  $ and multiplying both sides on the
left by $\mathcal{D}_{\alpha}^{\dag}(z)$ and comparing with (\ref{4.4}), we
get (\ref{4.8}). This completes the proof.
\end{proof}
\end{proposition}

The substitution of $\mathcal{S}_{\alpha}(z)=-iz\mathcal{K}_{\alpha}$ into
(\ref{4.5}) yields four possible and distinguishable solutions for
$\mathcal{K}_{\alpha}$: $\mathcal{Q}_{\alpha},\mathcal{Q}_{\alpha}^{\dag}$ and
$\mathcal{Q}_{\alpha}\pm\mathcal{Q}_{\alpha}^{\dag}$. The two-first cases will
be omitted to avoid ill-defined hermiticity condition imposed to
$\mathcal{K}_{\alpha}$, while the last case with a positive sign is that in
which we are interested in. Under these conditions, the displacement operator
(\ref{4.3}) can be expressed as:%
\begin{equation}
\mathcal{D}_{\alpha}(z)=\exp\left\{  z\left(  \mathcal{Q}_{\alpha}%
+\mathcal{Q}_{\alpha}^{\dag}\right)  \right\}  _{z=-z^{\ast}}=\exp\left\{
z\mathcal{Q}_{\alpha}-z^{\ast}\mathcal{Q}_{\alpha}^{\dag}\right\}  ,
\tag{4.9}\label{4.9}%
\end{equation}
and by inserting (\ref{4.1}) and (\ref{4.9}) into PDM CS (\ref{4.2}), we get%
\begin{equation}
\left\vert z;\alpha\right\rangle \sim U^{\alpha-1}(x)\exp\left\{  \sqrt
{2}z\mathcal{W}_{\alpha}(x)+\frac{1-2\alpha}{\sqrt{2}}zU^{\prime}(x)-%
{\displaystyle\int\nolimits_{{}}^{\mu(x)}}
\mathcal{W}_{\alpha}(\eta)d\mu(\eta)\right\}  . \tag{4.10}\label{4.10}%
\end{equation}

This expression is the general form of isospectral Hamiltonians CS endowed
with PDM for an arbitrary quantum system and already deduced in the case
$\alpha=1/2$ in \cite{29}. Let us now turn to the construction of PDM CS for
an accidental symmetry, $\overline{\left\vert z;\alpha\right\rangle }$, under
the $%
\mathbb{Z}
_{2}$ action and satisfying the eigenvalue equation $\mathcal{Q}_{1-\alpha
}\overline{\left\vert z;\alpha\right\rangle }=z\overline{\left\vert
z;\alpha\right\rangle }$. This procedure can easily be achieved in the same
manner it was performed in the exact SUSY symmetry or merely by performing the
substitution $\alpha\mapsto\overline{\alpha}=1-\alpha$ in (\ref{4.10}).
Finally, we get the relationship between both coherent states%
\begin{align}
\overline{\left\vert z;\alpha\right\rangle }  &  \sim U^{-\alpha}%
(x)\exp\left\{  \sqrt{2}z\mathcal{W}_{1-\alpha}(x)-\frac{1-2\alpha}{\sqrt{2}%
}zU^{\prime}(x)-%
{\displaystyle\int\nolimits_{{}}^{\mu(x)}}
\mathcal{W}_{1-\alpha}(\eta)d\mu(\eta)\right\} \nonumber\\
&  =U^{1-2\alpha}(x)\exp\left\{  -\sqrt{2}z\widetilde{\mathcal{W}}_{\alpha
}(x)-\sqrt{2}(1-2\alpha)zU^{\prime}(x)+%
{\displaystyle\int\nolimits_{{}}^{\mu(x)}}
\widetilde{\mathcal{W}}_{\alpha}(\eta)d\mu(\eta)\right\}  \left\vert
z;\alpha\right\rangle , \tag{4.11}\label{4.11}%
\end{align}
where $\widetilde{\mathcal{W}}_{\alpha}(x)=\mathcal{W}_{\alpha}(x)-\mathcal{W}%
_{1-\alpha}(x).$

On the other hand, the relevance of (\ref{3.13}) is clear in the sense that if
$\left\vert z;\alpha\right\rangle $ is PDM CS of $\mathcal{Q}_{\alpha}$, then
$\widehat{\mathcal{Z}}\left\vert z;\alpha\right\rangle $ is a new PDM CS of
$\mathcal{Q}_{1-\alpha}$ under the $%
\mathbb{Z}
_{2}$-transformation. Indeed by acting (\ref{3.13}) on $\left\vert
z;\alpha\right\rangle $, taking into account that $\mathcal{Q}_{\alpha
}\left\vert z;\alpha\right\rangle =z\left\vert z;\alpha\right\rangle $, it
seems that $\overline{\left\vert z;\alpha\right\rangle }=\widehat{\mathcal{Z}%
}\left\vert z;\alpha\right\rangle $ solves the new eigenvalue PDM CS equation,
where the action of $\widehat{\mathcal{Z}}$ on $\left\vert z;\alpha
\right\rangle $ is well established in (\ref{4.11}). However by comparing
(\ref{4.10}) to (\ref{4.11}), it is easy to verify that $\overline{\left\vert
z;1-\alpha\right\rangle }=\left\vert z;\alpha\right\rangle $, so that
$\overline{\left\vert z;\alpha\right\rangle }=\widehat{\mathcal{Z}}%
\overline{\left\vert z;1-\alpha\right\rangle }$. The same identity can be
deduced for the state $\left\vert z;\alpha\right\rangle $, i.e., $\left\vert
z;\alpha\right\rangle =\widehat{\mathcal{Z}}^{-1}\left\vert z;1-\alpha
\right\rangle $.

As we can see, the $%
\mathbb{Z}
_{2}$ action acts on PDM CS (\ref{4.10}) as a functional factor representing
the implementation of the transformation and affects (\ref{4.10}) as long as
$\alpha\neq1/2$ and $U(x)$ remains a function. Otherwise, (i.e., $\alpha
=1/2$), the $%
\mathbb{Z}
_{2}$ action is broken and both PDM CS in (\ref{4.11})\ are reduced to be the
same state.

In the following one may prove that PDM CS $\left\vert z;\alpha\right\rangle $
of (\ref{4.10}) minimize the generalized position-momentum uncertainty
relation (GUR). To prove this, let us calculate the variances, $\left(
\Delta\mathcal{W}_{\alpha}\right)  ^{2}$ and $\left(  \Delta\Pi_{\alpha
}\right)  ^{2}$, of the superpotential and generalized momentum operators
given by%
\begin{equation}
\widehat{\mathcal{W}}_{\alpha}(x)=\frac{1}{\sqrt{2}}\left(  \mathcal{Q}%
_{\alpha}+\mathcal{Q}_{\alpha}^{\dag}\right)  -\frac{1-2\alpha}{\sqrt{2}%
}U^{\prime}(x),\quad\text{and\quad}\widehat{\Pi}_{\alpha}(x)=-\frac{i}%
{\sqrt{2}}\left(  \mathcal{Q}_{\alpha}-\mathcal{Q}_{\alpha}^{\dag}\right)  ,
\tag{4.12}\label{4.12}%
\end{equation}
and deduced from (3.6). By definition, the variance of an operator, say
$\widehat{\Theta}$, is defined as: $\left(  \Delta\Theta\right)
^{2}=\left\langle z;\alpha\left\vert \widehat{\Theta}^{2}\right\vert
z;\alpha\right\rangle -\left\langle z;\alpha\left\vert \widehat{\Theta
}\right\vert z;\alpha\right\rangle ^{2}$. Then with the help of the first
equation of (3.9), we find after some straightforward but lengthy calculation%
\begin{align}
\left\langle \mathcal{W}_{\alpha}\right\rangle  &  \equiv\left\langle
z;\alpha\left\vert \widehat{\mathcal{W}}_{\alpha}\right\vert z;\alpha
\right\rangle =-\frac{1-2\alpha}{2}U^{\prime}(x),\tag{4.13a}\label{4.13a}\\
\left\langle \Pi_{\alpha}\right\rangle  &  \equiv\left\langle z;\alpha
\left\vert \widehat{\Pi}_{\alpha}\right\vert z;\alpha\right\rangle =-i\sqrt
{2}z,\tag{4.13b}\label{4.13b}\\
\left\langle \mathcal{W}_{\alpha}^{2}\right\rangle  &  \equiv\left\langle
z;\alpha\left\vert \widehat{\mathcal{W}}_{\alpha}^{2}\right\vert
z;\alpha\right\rangle =-\frac{1}{2}U(x)\mathcal{W}_{\alpha}^{\prime
}(x)-(1-2\alpha)U^{\prime}(x)\mathcal{W}_{\alpha}(x)\nonumber\\
&  +\frac{1-2\alpha}{2}U(x)U^{\prime\prime}(x)-\frac{(1-2\alpha)^{2}}%
{4}U^{\prime2}(x),\tag{4.13c}\label{4.13c}\\
\left\langle \Pi_{\alpha}^{2}\right\rangle  &  \equiv\left\langle
z;\alpha\left\vert \widehat{\Pi}_{\alpha}^{2}\right\vert z;\alpha\right\rangle
=-2z^{2}+\frac{1}{2}U(x)\mathcal{W}_{\alpha}^{\prime}(x)+\frac{1-2\alpha}%
{4}U(x)U^{\prime\prime}(x), \tag{4.13d}\label{4.13d}%
\end{align}
keeping in mind that $z=-z^{\ast}$. Thus the use of definition of the variance
given above provides the product%
\begin{equation}
\left(  \Delta\mathcal{W}_{\alpha}\right)  ^{2}\cdot\left(  \Delta\Pi_{\alpha
}\right)  ^{2}=\frac{1}{4}\left(  \left\langle z;\alpha\left\vert \left[
\mathcal{Q}_{\alpha},\mathcal{Q}_{\alpha}^{\dag}\right]  \right\vert
z;\alpha\right\rangle \right)  ^{2}-\frac{1}{2}\mathcal{R}_{\alpha
}(x)\left\langle z;\alpha\left\vert \left[  \mathcal{Q}_{\alpha}%
,\mathcal{Q}_{\alpha}^{\dag}\right]  \right\vert z;\alpha\right\rangle ,
\tag{4.14}\label{4.14}%
\end{equation}
where the function $\mathcal{R}_{\alpha}(x)$ is defined as%
\begin{equation}
\mathcal{R}_{\alpha}(x)=(1-2\alpha)U^{\prime}(x)\left(  \mathcal{W}_{\alpha
}(x)+\frac{1-2\alpha}{2}U^{\prime}(x)\right)  . \tag{4.15}\label{4.15}%
\end{equation}

As we can see a deeper insights are necessary on the function $\mathcal{R}%
_{\alpha}(x)$, if we are interested to saturate and minimize (\ref{4.14}).

\subparagraph{Saturation of (\ref{4.14}).}

Using (\ref{4.15}), it is worth noting that PDM CS $\left\vert z;\alpha
\right\rangle $ saturate GUR if and only if $\mathcal{R}_{\alpha}(x)=0$. This
latter leads us to consider three requirements: (\textbf{R1}) $\alpha=1/2$,
(\textbf{R2}) $U^{\prime}(x)=0$, and/or (\textbf{R3}) $\mathcal{W}_{\alpha
}(x)=-(1-2\alpha)U^{\prime}(x)/2.$ It is easy to convince ourselves that
(\textbf{R2}) must be avoided since in this case the quantum system loses its
PDM features. However if the restriction (\textbf{R1}) is satisfied, then
(\textbf{R3}) yields $\mathcal{W}_{1/2}(x)=0$ which corresponds to a free
particle in (\ref{3.7a}). This is in contrast with our study since the case
$\alpha=1/2$ has its own PDM potential and can not be zero. Thus in order to
avoid this ambiguity, we suggest that $\mathcal{W}_{\alpha}(x)\neq
-(1-2\alpha)U^{\prime}(x)/2$ simultaneously with $\alpha=1/2$. On the other
hand, taking into account only the restriction (\textbf{R3}) we get from (3.7)%
\begin{equation}
\mathcal{V}_{\text{eff}}(x)=-\frac{1}{4}U(x)U^{\prime\prime}(x)-\frac{1}%
{8}U^{\prime2}(x)+\epsilon,\tag{4.16}\label{4.16}%
\end{equation}
which is independent of $\alpha.$ We conclude that\textit{ the saturation of
PDM CS (\ref{4.10}) is possible if and only if:}

\begin{itemize}
\item $\alpha=1/2$ and $\mathcal{W}_{\alpha}(x)\neq-(1-2\alpha)U^{\prime
}(x)/2$,

\item $\alpha\neq1/2$ and $\mathcal{W}_{\alpha}(x)=-(1-2\alpha)U^{\prime
}(x)/2$, leading to the effective potential (\ref{4.16}).
\end{itemize}

\subparagraph{Minimization of (\ref{4.14}).}

On the other context, it is also simple to verify that PDM CS, $\left\vert
z;\alpha\right\rangle $, minimize GUR as follows%
\begin{equation}
\left(  \Delta\mathcal{W}_{\alpha}\right)  ^{2}\cdot\left(  \Delta\Pi_{\alpha
}\right)  ^{2}\geq\frac{1}{4}\left(  \left\langle z;\alpha\left\vert \left[
\mathcal{Q}_{\alpha},\mathcal{Q}_{\alpha}^{\dag}\right]  \right\vert
z;\alpha\right\rangle \right)  ^{2},\tag{4.17}\label{4.17}%
\end{equation}
if and only if $\mathcal{R}_{\alpha}(x)<0$. This restriction leads us to
distinguish between two possible cases from (\ref{4.15}) as follows (with
$0\leq\alpha\leq1$):

\begin{description}
\item[(C1)] If $(1-2\alpha)U^{\prime}(x)<0$, then $\mathcal{W}_{\alpha
}(x)+(1-2\alpha)U^{\prime}(x)/2>0,$

\item[(C2)] If $(1-2\alpha)U^{\prime}(x)>0$, then $\mathcal{W}_{\alpha
}(x)+(1-2\alpha)U^{\prime}(x)/2<0.$
\end{description}

The first condition of (\textbf{C1}) gives either $-1\leq1-2\alpha<0$ with
$U^{\prime}(x)>0$, or $0<1-2\alpha\leq1$ for $U^{\prime}(x)<0$. Then, in both
situations, the second condition yields $\mathcal{W}_{\alpha}(x)>0$. The same
analysis is made for (\textbf{C2}), which gives either $-1\leq1-2\alpha<0$
with $U^{\prime}(x)<0$, or $0<1-2\alpha\leq1$ for $U^{\prime}(x)>0$, while the
second condition of this case yields $\mathcal{W}_{\alpha}(x)<0$. We conclude
here that \textit{the sign of the superpotential }$\mathcal{W}_{\alpha}%
(x)$\textit{ is entirely controlled by the minimization of GUR obtained
fr\textit{om (\ref{4.10})}}.

\section{Concluding remarks}

We have extended the ideas given in \cite{33} for an arbitrary quantum system
of isospectral family of Hamiltonians endowed with PDM and constructed their
PDM CS through the unitary transformation.

In this paper, our main aim is to emphasize the greatest importance of
introducing the unitary transformation and its consequence for solving the
ordering problem. For instances, we have shown that the profile of the mass
function $m(x)$ (resp. $U(x)$) is subjected to \textit{singularities} (resp.
\textit{zeros}). In this way, we have been able to define a special set of the
SUSY parameter $\alpha$ which distinguishes different choices of the kinetic
operator $\widehat{T}_{\alpha}$ and some of the mostly used forms are reported
in table 1. We have also observed the occurrence of an accidental symmetry
under the $%
\mathbb{Z}
_{2}$-transformation if the special value $\alpha=1/2$ is the unique fixed
point of this transformation and proved that this point is the most suitable
choice under this set. This last comment may be the most subtle idea in our
paper because it suggests that the $%
\mathbb{Z}
_{2}$ action is rather simple for explaining the origin of the very special
ordering $\widehat{T}_{1/2}=\frac{1}{2}\sqrt{U(x)}\widehat{P}U(x)\widehat
{P}\sqrt{U(x)}$, due to the special feature of the point $\alpha=1/2$. We have
seen that there is another remarkable property concerning the fixed point,
that is the operator $\widehat{\mathcal{Z}}$ which implements the $%
\mathbb{Z}
_{2}$ action acts as $\mathcal{Q}_{1/2}$ if $\alpha=1/2$, however it plays the
role of an intertwiner operator between $\mathcal{Q}_{\alpha}$ and
$\mathcal{Q}_{1-\alpha}$ for each $\alpha\neq1/2$ of the set $\mathfrak{S}%
_{\alpha_{n}}$.

We have constructed PDM CS in the Klauder-Perelomov's sense through the exact
SUSY symmetry and proved that the $%
\mathbb{Z}
_{2}$ action acts on them as a functional factor as long as $\alpha\neq1/2$.
Finally we have also shown that the fixed point has another remarkable effect;
its associated PDM CS saturate the generalized position-momentum uncertainty
relation, while this latter is minimized for $\alpha\neq1/2$. A quite
important result of this minimization is that the sign of the superpotential
$\mathcal{W}_{\alpha}(x)$ is entirely determined.


\begin{thebibliography}{99}                                                                                               %


\bibitem[1]{1}B. Mielnik, J. Math. Phys. \textbf{25,} 3387 (1984).

\bibitem[2]{2}M. M. Nieto, Phys. Lett. \textbf{145B}, 208 (1984).

\bibitem[3]{3}S. H. Dong, \textit{Factorization Method in Quantum Mechanics}
(The Netherlands, Springer, 2007) and references therein.

\bibitem[4]{4}K. Chadan and P.\ C.\ Sabatier, \textit{Inverse Problems in
Quantum Scattering Theory} (Springer, Berlin, 1977).

\bibitem[5]{5}F. Cooper, A. Khare and U. Sukhatme, Phys. Rep. \textbf{251,}
267 (1995).

\bibitem[6]{6}F. Iachello, \textit{Lie Algebra and Applications}. Lect. Notes
Phys. 708 (Springer, Berlin, 2006).

\bibitem[7]{7}R. Bravo and M. S. Plyushchay, Phys. Rev. D \textbf{93,} 105023
(2016) and references therein.

\bibitem[8]{8}J.-M. L\'{e}vy-Lebland, Phys. Rev. A \textbf{52,} 1845 (1995);
Eur. J. Phys. \textbf{13,} 215 (1992).

\bibitem[9]{9}B. Roy and P. Roy, J. Phys. A: Math. Gen. \textbf{35,} 3961 (2002).

\bibitem[10]{10}R. Ko\c{c} and M. Koka, J. Phys. A: Math. Gen. \textbf{36},
8105 (2003).

\bibitem[11]{11}B. Bagchi, B. Banerjee, C. Quesne and V. M. Tkachuk, J. Phys.
A: Math. Gen. \textbf{38,} 2929 (2005).

\bibitem[12]{12}A. Gunguly and L. M. Nieto, J. Phys. A: Math. Theor.
\textbf{40,} 7265 (2007).

\bibitem[13]{13}A. A. Suzko and A. Schulze-Halberg, Phys. Lett. A
\textbf{372}, 5865 (2008).

\bibitem[14]{14}S.-A. Yahiaoui and M. Bentaiba, Int. J. Theor. Phys.
\textbf{48,} 315 (2009).

\bibitem[15]{15}B. Midya and B. Roy, Phys. Lett. A \textbf{373,} 4117 (2009).

\bibitem[16]{16}G. L\'{e}vai and O. \"{O}zer, J. Math. Phys. \textbf{51},
092103 (2010).

\bibitem[17]{17}S. H. Mazharimousavi, Phys. Rev. A \textbf{85,} 034102 (2012).

\bibitem[18]{18}P. Ring and P. Schuck, \textit{The Nuclear Many Body Problems}
(Springer, New York, 1980), p. 211.

\bibitem[19]{19}C. Quesne and V. M. Tkachuk, J. Phys. A: Math. Gen.
\textbf{37,} 4267 (2004).

\bibitem[20]{20}L. Jiang, L.-Z. Yi and C.-S. Jia, Phys. Lett. A \textbf{345,}
279 (2005).

\bibitem[21]{21}O. Mustafa and S. H. Mazharimousavi, J. Phys. A: Math. Theor.
\textbf{41,} 244020 (2008) and references therein.

\bibitem[22]{22}P. Harrison, \textit{Quantum Wells, Wires and Dots.
Theoretical and Computational Physics of Semiconductor Nanostructures} (John
Wiley \& Sons, LTD, 2005).

\bibitem[23]{23}G. Bastard, \textit{Wave Mechanics Applied to Semiconductor
Heterostructures} (Les Ulis: \'{E}ditions de physique, 1998).

\bibitem[24]{24}O. von Roos, Phys. Rev. B \textbf{27,} 7547 (1983).

\bibitem[25]{25}E. Schr\"{o}dinger, Naturwiss \textbf{14}, 664 (1926).

\bibitem[26]{26}R. J. Glauber, Phys. Rev. \textbf{131}, 2766 (1963).

\bibitem[27]{27}D. J. Fern\'{a}ndez, V. Hussin and L. M. Nieto, J. Phys. A:
Math. Gen. \textbf{27,} 3547 (1994).

\bibitem[28]{28}J.-P. Gazeau, \textit{Coherent States in quantum Mechanics
}(Wiley-VCH, Weinheim, 2009).

\bibitem[29]{29}S.-A. Yahiaoui and M. Bentaiba, J. Phys. A: Math. Theor.
\textbf{47,} 025301 (2014); J. Phys. A: Math. Theor. \textbf{45,} 444034 (2012).

\bibitem[30]{30}V. C. Ruby and M. Senthilvelan, J. Math. Phys. \textbf{51,}
52106 (2010).

\bibitem[31]{31}S. Cruz y Cruz and O. Rosas-Ortiz, J. Phys. A: Math. Theor.
\textbf{42,} 185205 (2009); Int. J. Theor. Phys. \textbf{50,} 2201 (2011).

\bibitem[32]{32}N. Amir and S. Iqbal, J. Math. Phys. \textbf{56}, 062108
(2015); Commun. Theor. Phys. \textbf{66,} \ 41 (2016).

\bibitem[33]{33}M. S. Kumar and A. Khare, Phys. Lett. A \textbf{217,} 73 (1996).

\bibitem[34]{34}R. A. Morrow and K. R. Brownstein, Phys. Rev. B \textbf{30,}
678 (1984).

\bibitem[35]{35}Q. J. Zhu and H. Kroemer, Phys. Rev. B \textbf{27,} 3519 (1983).

\bibitem[36]{36}D. J. BenDaniel and C. B. Duke, Phys. Rev. B \textbf{152}, 683 (1966).

\bibitem[37]{37}M.-A. Fiset and V. Hussin, J. Phys.: Conf. Ser. \textbf{624,} 012016\ (2015)
\end{thebibliography}
\end{document}